\documentclass{article}
\usepackage{graphicx}
\usepackage{amsmath}

\begin{document}
\begin{center}
{\bf \Large How pairs of partners emerge in an initially fully connected society}\\[5mm] 
{\large J.~Karpi\'nska, K.~Malarz$^*$ and K.~Ku{\l}akowski$^\dag$}\\[3mm]
{\em Department of Applied Computer Science, 
Faculty of Physics and Nuclear Techniques, 
AGH University of Science and Technology\\
al. Mickiewicza 30, PL-30059 Krak\'ow, Poland}

\bigskip

e-mail: $^*${\tt malarz@agh.edu.pl}, $^\dag${\tt kulakowski@novell.ftj.agh.edu.pl}

\bigskip

\today
\end{center}

\begin{abstract}
A social group is represented by a graph, where each pair of nodes is
connected by two oppositely directed links. At the beginning, a given amount
$p(i)$ of resources is assigned randomly to each node $i$. Also, each link
$r(i,j)$ is initially represented by a random positive value, which means the
percentage of resources of node $i$ which is offered to node $j$. Initially
then, the graph is fully connected, i.e. all non-diagonal matrix elements
$r(i,j)$ are different from zero. During the simulation, the amounts of
resources $p(i)$ change according to the balance equation. Also, nodes
reorganise their activity with time, going to give more resources to those
which give them more. This is the rule of varying the coefficients $r(i,j)$.
The result is that after some transient time, only some pairs $(m,n)$ of nodes
survive with non-zero $p(m)$ and $p(n)$, each pair with symmetric and positive
$r(m,n)=r(n,m)$. Other coefficients $r(m,i\ne n)$ vanish. Unpaired nodes remain with no
resources, i.e. their $p(i)=0$, and they cease to be active, as they have
nothing to offer. The percentage of survivors (i.e. those with with $p(i)$
positive) increases with the velocity of varying the numbers $r(i,j)$, and it
slightly decreases with the size of the group. The picture and the results can
be interpreted as a description of a social algorithm leading to marriages.
\end{abstract}



\section{Introduction}
In the graph theory, several interesting problems are formulated in terms of
marriages \cite{wilson}.
Example giving, the famous Hall marriage theorem precises the number of links between a given number of nodes in a bipartite graph which are necessary if all these nodes are to be paired.
The interpretation is that the nodes represent persons of given sex, and marriages are possible only between persons which are linked: they know each other.

However, almost all of us know that these links are not rigid.
It is possible to imagine that somebody marries an almost unknown person, just because of her/his appealing view.
The reader can find many case studies in numerous books (e.g. \cite{tolstoy}) and journals even if they are not entirely devoted to statistical phenomena or computer simulations.
Here we are inspired by the considerations on the dynamics of marital interaction, described in \cite{gott,murr}.
There, the interaction between partners is presented as a kind of game, where attracting or repulsing behaviours of both sides lead either to a final agreement, or to a developing conflict which must end by the separation.

According to our perspective, the same pattern is present at a social scale.
Each member of a given group sends simultaneous signals to numerous other
members. Those who, for some individual reasons, feel more attracted by a
given person, orientate their attention at least partially and send to this
person in turn signals more expressive. This is a positive feedback. On the
contrary, those who remark a decrease of their partner's interest try also
to reorient their efforts to find somebody more promising, where more
rewards can be expected. In this way, the initially multiple contacts are
polarised.

The results presented here demonstrate that in fact this kind of feedback is
entirely sufficient for a complete transformation of the whole structure.
Initially everybody gives some attention to everybody. Keeping the graph
terminology, initially links exist between each two members. At the final
state, however, only some pairs of links exist between members of separate
pairs.

Let us add that multiple examples in literature and life teach us that the
ability to devote attention to other people is gradually used up (see, for example \cite{rostand}).
After some years, we wake up with greater emotional distance to the surrounding
world. If we are already married, memory on our past feelings maintains our
optimism and confidence. Remaining single, we have no motivation to search
for a partner any more. Discussion of social or psychological details of this
effect obviously exceeds the frames of any text devoted to a computation.

Looking for a mathematical description of this picture, we consider the
social group as a set of $N$ nodes $i=1,\cdots,N$.
Their emotional resources are given by $p(i)$.
The flow of these resources from $i$ to $j$ is described by a matrix element $r(i,j)$.
In the graph terminology, a directed link from $i$ to $j$ exists if $r(i,j)$ is positive.

In the next section, equations are given which are designed to model the time
evolution of the vector $p(i=1,\cdots,N)$ and of the matrix $[r(i,j)]$.
Subsequent sections contain the results of the simulation and the conclusions.

\section{The model}
The emotional resources of $i^{\text{th}}$ unit is defined as $p(i)\ge 0$. It evolves according to the following rule:
\begin{equation}
\frac{dp(i)}{dt}=N^2-\left[\sum_{j=1}^Np(j)\right]^2+\sum_{j=1}^N\left[r(j,i)p(j)-r(i,j)p(i)\right].
\label{eq1}
\end{equation}

First two terms on the right are introduced for a computational convenience to
keep $p(i)$ summed over $i$ determined and finite. They are equivalent to a
kind of Verhulst term in population models \cite{murr}. Although they have no
direct importance for our subject here, they could be of interest for
a reformulation our model to capture some economical phenomena \cite{wars}.
The third term in square brackets represents the exchange of emotional resources $p$ between $i$ and $j$.
The summation over $j$ gives the total flow of the resources to and from the $i^{\text{th}}$ node, weighted with $r(i,j)$.

Let us denote $\sum_ip(i)\equiv S$.
Summing up Eq. \eqref{eq1} over $i$, we get 
\begin{equation}
\frac{dS}{dt}=N^3-NS^2+\text{zero}, 
\label{eq2}
\end{equation}
because the remaining terms cancel each other.
Then, $S=N$ is the only stable point.
Having determined $S$ --- for each matrix $[r(i,j)]$ --- we have one and only one solution $p(i=1,\cdots,N)$, where simultaneously $\forall 1\le i\le N: dp(i)/dt=0$ and $dS/dt=0$.

We assume that the matrix $[r(i,j)]$ evolves according to the following rule:
\begin{equation}
\frac{dr(i,j)}{dt}=\alpha \left( r(j,i)p(j)-\frac{\sum_k r(k,i)p(k)}{N-1} \right).
\label{eq3}
\end{equation}
This rule reflects the tendency to offer more to a partner which offers more.
Note that the actual amount of resources transferred from $i$ to $j$ is the
product of $r(i,j)$, what is `intention' to give, and of $p(i)$, what is the
amount at the $i$'s disposition. The coefficient $\alpha$ is a common velocity
of reorientation, while the velocity of varying $p(i)$ is unity. As it is
demonstrated below, $\alpha$ appears to be an important measure of an ability to find a partner.

It is easy to check that the evolution preserves $N$ constants of motion
$\sum_j r(i,j)$. Additional conditions are that $p(i)\ge 0$. Once
$p(i)$ crosses zero, it is forced by the computer code to remain zero till the
end of simulation. At these time moments, the above mentioned constants of
motion may be disturbed. To omit the effect, for each $i$ we normalise
$\sum_j r(i,j)$ to be equal to unity at each step of the simulation.

We need to solve numerically $N^2$
differential equations. The matrix $[r(i,j)]$ has zeros at its
main diagonal. The additional number of $N$ equations for the $p$'s cancels
with the number of $N$ above mentioned constants of motion.

\section{Results}  
We solve numerically Eqs. \eqref{eq1} and \eqref{eq3} with randomly chosen initial conditions $0\le p(i)\le 1$ and $0\le r(i,j)\le 1$ for all $1\le i,j\le N$.
The simulation takes typically $N_{iter}=10^7$ steps and time step is set to $\Delta t=10^{-4}$ what ensures numerical stability of the solutions.
The results are averaged over $N_{run}=50$, 20 and 10 societies for $N=10$, 20 and 50, respectively.

For the calculation time long enough, we get a specific distribution of $p(i)$.
For some nodes $k$, $p(k)=0$.
For remaining nodes, the only non-zero matrix elements $r(i,j)$ are symmetric, i.e. $r(i,j)=r(j,i)$ and $p(i)=p(j)$.
This means, that the nodes $(i,j)$ form a pair.
Links from the pair members to the other partners vanish, i.e. $r(i,m\ne j)=0$ and $r(n\ne j,i)=0$.

Fig. \ref{eq1} shows the probability $P$ of finding a partner depending on the parameter $\alpha$ for different values of $N$.
Two examples of the time evolution of the individuals' amount of resources $p(i)$ for $N=10$ and two different values of $\alpha$ are presented in Fig. \ref{fig2}.

\begin{figure}
\includegraphics[width=.9\textwidth]{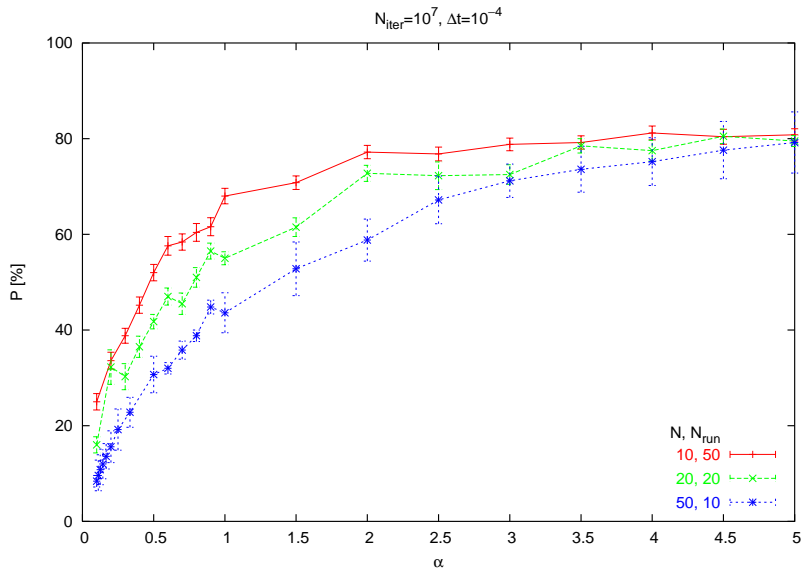}
\caption{The probability $P$ to find a partner, as dependent on the parameter $\alpha$ for different society sizes $N$.}
\label{fig1}
\end{figure} 

\begin{figure}
(a) \includegraphics[width=.9\textwidth]{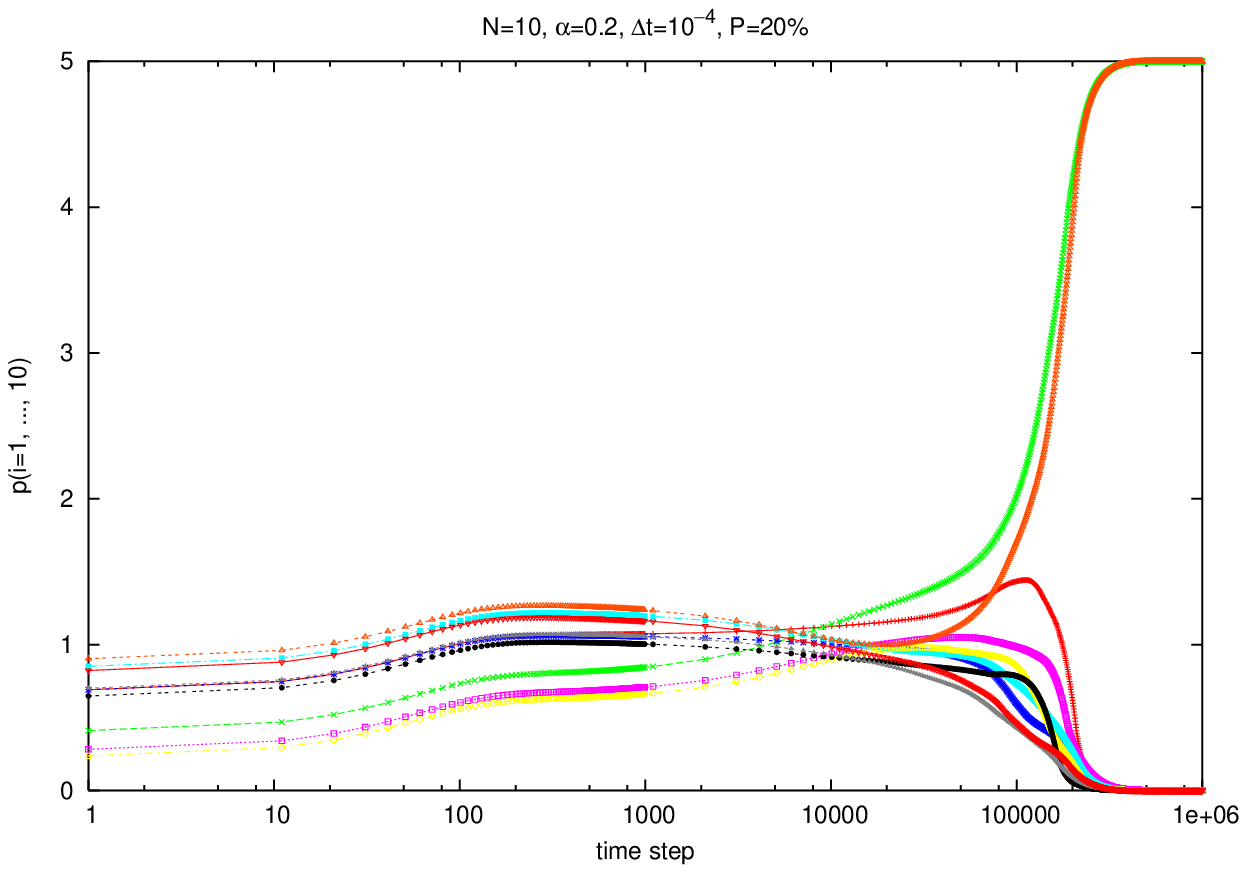}\\
(b) \includegraphics[width=.9\textwidth]{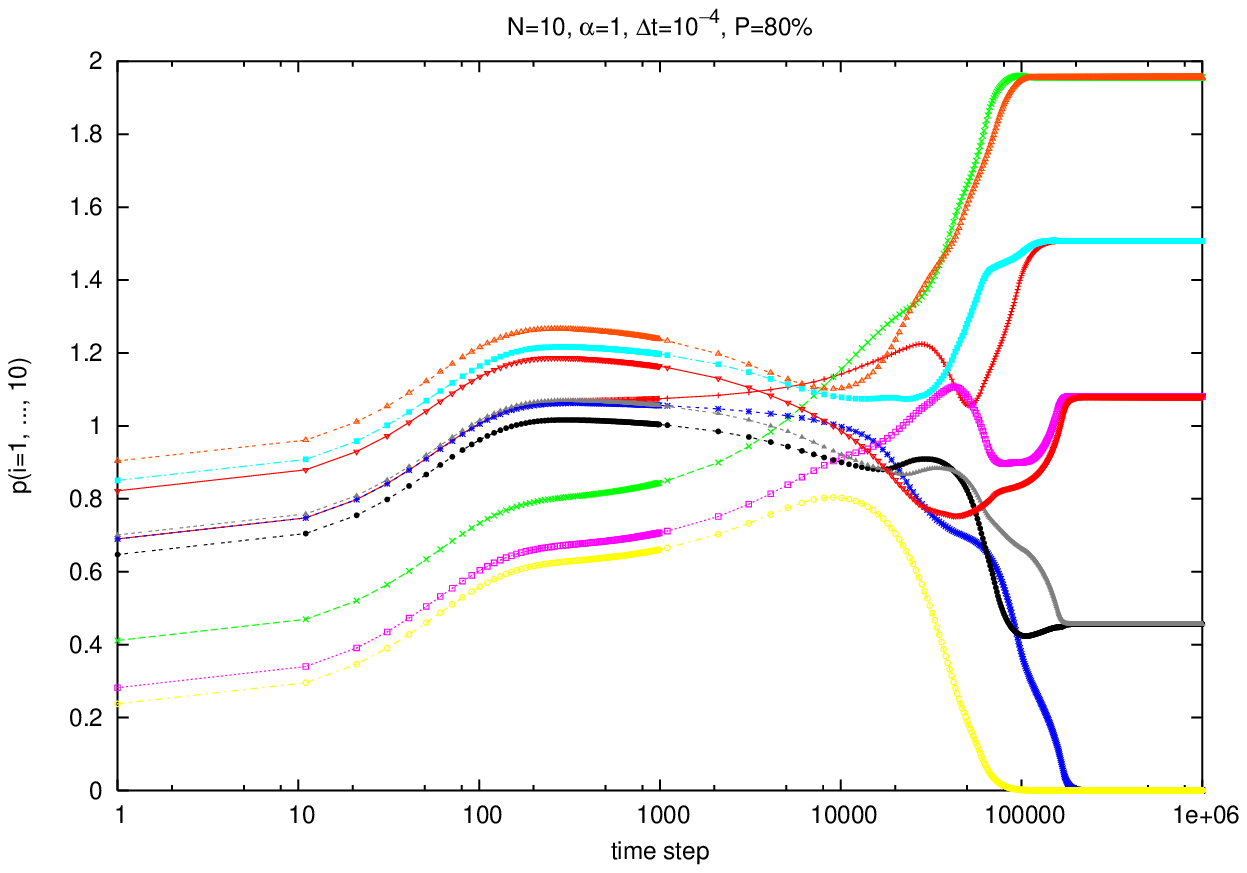}
\caption{The time evolution of the amount of individual resources $p(1\le i\le N)$ for small society ($N=10$) for different values of (a) $\alpha=0.2$ (b) $\alpha=1.0$ parameter with the same initial conditions.}
\label{fig2}
\end{figure} 

Additionally, we have checked that for relatively quick reorientation ($\alpha=3$) and very small population ($N=4$) roughly 25\% of $N_{run}=10^4$ investigated societies evolved to the state with one pair only and 75\% of societies finished with two pairs. 
No `trios' or `quartets' are observed.

\section{Conclusions}
Direct results of our simulations point out that the probability $P$ to find a suitable partner depends on two parameters: $\alpha $ and $N$. It is obvious that $P$ increases with $\alpha$, because the direct meaning of the latter is the
velocity of reorientation.
Strictly, $\alpha $ is the ratio of the velocity of evolution of the matrix elements $r(i,j)$ to the velocity of evolution of the elements $p(i)$.
What was not expected is the change of the plot slope near $\alpha=1$ (Fig. \ref{fig1}). One could conclude that it is sufficient to change partners only as quick as we fed them with our sympathy, if these velocities could be measured at the same scale.

The reduction of $P$ with $N$ reflects an obvious fact that it is more
difficult to find a partner in a larger pool. It means that the group members
are more likely to remain unmarried when $N$ is large. This is due to the
finite time which they have to find a partner, not involved in another
relationship. The timescale is determined here by the time of reduction of a
given $p(i)$ due to the lack of feedback. Once our emotional resources are
fed up, we have no possibility to overcome bonds of another pair, which are
already tight.

Here, the same value of the parameter $\alpha $ is assigned to all group
members. This is an obvious simplification. It is clear that in actual
societies one can observe a wide distribution of the parameter, which could be
termed `infidelity' or `ease to change partners' or `cunning' or `far-sight',
depending on the intention. Indirectly, our results confirm a simple
connection between this parameter and the probability of finding a partner.
Directly, our result deal with statistics, and not with individual
competition. For this purpose, the above simplification can be accepted.

In our opinion, most intriguing result is the observed lack of larger
clusters, e.g. triangles, members of which could support each other.
Despite the fact, that according to Eq. \eqref{eq3}, members are promoted the input
of which is just above the average, we have never observed a triangle even for
simulations for $N=4$. We must conclude that triangles are unlikely.

A direct conclusion from Fig.~\ref{fig1} seems to be that in any case, the percentage of never-married people is about twenty percent.
In fact, some statistical data confirm that this evaluation is realistic: in Ref. \cite{us1997} we find 23.5~\% for men and 19.2~\% for women.
However, these data cannot be taken without precaution.
The percentage of never-married people is known to increase with time \cite{us2001} and vary from place to place \cite{mass}.
On the other hand, our data are obtained from analysis of just one method of finding a partner, i.e. direct search and personal attracting.
This method can be considered as typical only in a modern society.
Some years ago, the society (family, church) was involved to match young people \cite{fidler}.
It is well possible, that this traditional method was more efficient.

Finally, we would like to broaden a little bit the field of applicability of
the results, not to insult the reader formed in the realistic world. We speak on
marriages only for simplicity, and in fact many other relationships can be put
instead, realistic or virtual. For our results, even the hypothesis of sex is
not necessary.

\section*{Acknowledgement}
The part of calculations was carried out in ACK\---CY\-FRO\-NET\---AGH.
The machine time on HP Integrity Superdome is financed by Ministry of Science and 
Information Technology in Poland under grant No. MNiI/\-HP\_I\_SD/\-AGH/\-002/\-2004.



\begin{thebibliography}{88}
\bibitem{wilson} R.~J.~Wilson, 
{\it Introduction to Graph Theory}, 
Addison Wesley Longman Ltd., London 1996.

\bibitem{tolstoy} L.~Tolstoy, 
{\it Anna Karenina},
{\tt http://www.gutenberg.net/\-etext/\-1399}.

\bibitem{gott} J.~M.~Gottman, 
{\it Marital Interaction: Experimental Investigations}, 
Academic Press, New York 1979.

\bibitem{murr} J.~D.~Murray, 
{\it Mathematical Biology. I. An Introduction,}
Springer-Verlag, Berlin 2002. 

\bibitem{rostand}
E.~Rostand, 
{\it Cyrano de Bergerac},
{\tt http://www.gutenberg.net/\-etext/\-1254}.

\bibitem{wars} K.~Malarz and K.~Ku{\l}akowski, 
in {\it From Quanta to Societies}, W. Klonowski (Ed.), 
Proc. of Euroattractor 2001, PABST, Lengerich 2003, p. 192 ({\tt cond-mat/0104487}).

\bibitem{us1997} {\it March 1997 Current Population Survey},
U.S. Census Bureau
({\tt http://www.unmarriedamerica.org/workplace/statistics.htm}).

\bibitem{us2001} {\it Statistical Abstracts of the United States 2001},
U.S. Bureau of the Census
({\tt http://www.infoplease.com/ipaA0763219.html}).

\bibitem{mass} {\tt http://www.massstats.com/demographics/map.asp}.

\bibitem{fidler} S.~Aleichem,
{\it Tevye Daughters: Collected Stories of Sholom Aleichem},
JBH of Peconic Inc., 1999 --- better known as {\it Fiddler on the Roof} by J.~Bock (music), S.~Harnick (lyrics) and J.~Stein (book).

\end{thebibliography}
\end{document}